\documentclass{optica-article}

\journal{opticajournal} 

\articletype{Research Article}

\usepackage{lineno}

\begin{document}

\title{Room-temperature high-average-power strong-field terahertz source based on industrial high-repetition-rate femtosecond laser}

\author{Deyin Kong,\authormark{1,2,3} Yichen Su,\authormark{4} Cheng Song,\authormark{4} and Xiaojun Wu,\authormark{1,2,3,*}}

\address{\authormark{1}Hangzhou International Innovation Institute, Beihang University, 166 Shuanghongqiao Street, Pingyao Town, Yuhang District, Hangzhou 311115, China\\
\authormark{2}School of Electronic and Information Engineering, Beihang University, 37 Xueyuan Road, Haidian District, Beijing 100191, China\\
\authormark{3}Zhangjiang Laboratory, 100 Haike Road, Shanghai 201210, China\\
\authormark{4}Key Laboratory of Advanced Materials (MOE), School of Materials Science and Engineering, 
Tsinghua University, 30 Shuangqing Road, Haidian District, Beijing 100084, China.}

\email{\authormark{*}xiaojunwu@buaa.edu.cn} 
{\raggedright\footnotesize\itshape{Preprint submitted to Photonics Research}}

\begin{abstract*} 
Free-space strong-field terahertz (THz) pulses, generated via optical rectification of femtosecond lasers in nonlinear crystals, are pivotal in various applications. However, conventional Ti:sapphire lasers struggle to produce high-average-power THz due to their limited output power. While kilowatt ytterbium lasers are increasingly adopted, their application in THz generation faces challenges: low optical-to-THz conversion efficiency (attributed to long pulse durations and low energy) and crystal damage under high pumping power. Here, we report a high-average-power strong-field THz source using a lithium niobate crystal pumped by a 1030-nm, 570-fs, 1-mJ, 50-kHz ytterbium femtosecond laser with tilted pulse front pumping (TPFP). By systematically optimizing TPFP implementations and comparing grating- and echelon-type configurations, we achieve a THz source with 64.5~mW average power at 42-W, 50-kHz pumping, and a focused peak electric field of 525 kV/cm at 0.83-mJ, 1-kHz operation. Additionally, we observe Zeeman torque signals in cobalt-iron ferromagnetic nanofilms. This high-repetition-rate, high-average-power THz system, combined with its potential capabilities in high signal-to-noise spectroscopy and imaging, promises transformative impacts in quantum matter manipulation, non-destructive testing, and biomedicine.

\end{abstract*}

\section{Introduction}
Free-space strong-field terahertz (THz) technology has revolutionized materials science and condensed matter physics by leveraging ultrashort pulse durations and intense electric/magnetic fields to manipulate matter at the atomic and molecular scales, such as inducing material nonlinearity, coupling magnons, deflecting magnetization, and exciting vibrational phonon modes\cite{luNonlinearOpticalPhysics2024,hoffmannTerahertzKerrEffect2009,salikhovCouplingTerahertzLight2023,chekhovBroadbandSpintronicDetection2023,zhangProbingUltrafastDynamics2021,bayerTerahertzLightMatter2017}. Besides, these capabilities also hold significant promise for electron acceleration and biomedical applications, as recently highlighted by advances in ultrafast science and biophotonics\cite{zhangSegmentedTerahertzElectron2018,nanniTerahertzdrivenLinearElectron2015,liuRecentAdvancesResearch2023}. However, the critical bottleneck restricting the advancement of strong-field THz science and technological applications lies in the lack of high-efficiency, high-beam-quality, and high-stability THz sources, which remain challenging to realize due to the fundamental trade-offs between power scaling and pumping fluence. Femtosecond laser-matter interaction has emerged as a cornerstone technique for generating strong-field THz electromagnetic pulses, including but not limited to optical rectification\cite{tothTiltedPulseFront2023}, air plasma\cite{koulouklidisObservationExtremelyEfficient2020}, and spintronic emitters\cite{seifertEfficientMetallicSpintronic2016}. Among these approaches, optical rectification in lithium niobate crystals pumped by Ti:sapphire femtosecond lasers has emerged as a fundamental technology owing to its high pumping energy and stability. Lithium niobate crystals have been widely used in various applications due to their high nonlinear coefficient and damage threshold\cite{xieBroadbandMillimeterwaveFrequency2025,boesLithiumNiobatePhotonics2023}. Decades of intensive research have culminated in the realization of strong-field THz sources generating single-pulse energy exceeding 10 mJ, achieved through pumping by 1-Hz Ti:sapphire lasers\cite{wuGeneration139mJTerahertz2023}. However, advanced precision measurements and commercial implementations increasingly demand strong-field THz sources with high average power, high repetition rate, and high beam quality, as exemplified by applications such as strong-field THz-coupled angle-resolved photoemission spectroscopy (ARPES)\cite{reimannSubcycleObservationLightwavedriven2018}, scanning near-field optical microscopy (SNOM)\cite{cockerNanoscaleTerahertzScanning2021}, and scanning tunnelling microscope (STM)\cite{cockerUltrafastTerahertzScanning2013}. Moreover, high-average-power THz sources offer signal intensities orders of magnitude stronger than weak-field THz time-domain spectrometers, enabling high-quality THz imaging\cite{liHighthroughputTerahertzImaging2023} and computed tomography\cite{fosodederHighlyAccurateTHzCT2022} while unlocking applications previously constrained by limited photon flux. These advancements underscore the pressing need for next-generation high-average-power strong-field THz systems to realize their full scientific and technological potential. 

In pursuit of this objective, recent advancements in high-average-power ytterbium (Yb) femtosecond lasers\cite{brauchHighpowerHighbrightnessSolidstate2022} have enabled the generation of high-average-power, strong-field THz radiation via lithium niobate crystal pumping, marking a paradigm shift from conventional Ti:sapphire laser systems. Notwithstanding latest breakthroughs in high-average-power THz generation via Yb laser pumping of lithium niobate crystals, including a record 1.3\% conversion efficiency\cite{guiramandNearoptimalIntensePowerful2022} and 643-mW average power\cite{vogelSinglecycle643MW2024}, this field remains constrained by fundamental trade-offs. Commercially available Yb lasers, while offering high average power, typically exhibit low pulse energy (<1 mJ) and long pulse durations (>200 fs) compared with those of Ti:sapphire lasers, limiting optical-to-THz conversion efficiency. Reducing the pumping laser beam size to enhance fluence improves optical rectification efficiency but risks crystal damage. Additionally, thermal effects from high-repetition-rate pumping further degrade efficiency, surpassing those observed in low-repetition-rate Ti:sapphire systems. These challenges necessitate innovative approaches to achieve efficient THz generation under low-energy, long-pulse conditions without compromising crystal integrity. A critical enabling technology is tilted pulse front pumping (TPFP), which synchronizes pumping laser and THz pulse fronts for velocity-matched optical rectification. Conventional grating-based TPFP\cite{heblingVelocityMatchingPulse2002} can achieve >99\% diffraction efficiency at Littrow angle, but introduces angular dispersion that broadens pulse durations. Conversely, echelon-mirror TPFP mitigates dispersion\cite{ofori-okaiTHzGenerationUsing2016} but suffers from low reflectivity, reducing THz output. Resolving this TPFP design dilemma, alongside thermal management and fluence optimization, is essential to realize high-performance THz sources capable of meeting the demands of next-generation applications.

In this study, we demonstrate a high-average-power strong-field THz source generating 64.5~mW average power via TPFP in a lithium niobate crystal using a 1030-nm Yb laser with 570-fs pulse duration and 0.83-mJ pulse energy at 50 kHz repetition rate. Despite the low pulse energy, this configuration achieves a focused peak electric field of 525 kV/cm at a 1 kHz repetition rate. Importantly, we systematically compare two TPFP implementations: a transmission grating operating at a Littrow angle and a reflection echelon mirror illuminated by the laser at normal incidence. While both techniques enable velocity matching for efficient THz generation, the grating-based approach outperforms the echelon design under long-pulse-duration, low-energy pumping conditions. This finding resolves the TPFP design dilemma in high-repetition-rate systems with \textasciitilde 500-fs pulse duration and <1 mJ pulse energy. To validate the capabilities of this system, we characterize Zeeman torque dynamics in cobalt-iron (CoFe) ferromagnetic thin films, achieving sub-picosecond temporal resolution in magnetization dynamic measurements. Collectively, this system provides a robust platform for exploring nonlinear optical phenomena, quantum state manipulation, and other spectroscopic imaging applications.

\section{Experimental setup}

\subsection{Overall system}
\begin{figure}[htbp]
\centering\includegraphics[width=\textwidth]{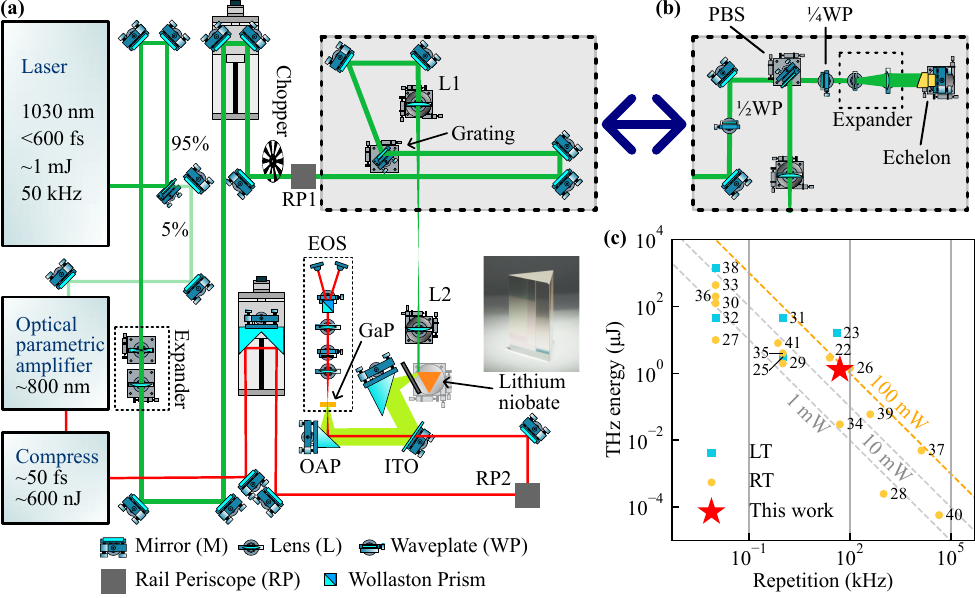}
\caption{The generation and detection experimental setup of high-average-power strong-field THz system. (a) Schematic diagram of the optical system. (b) The echelon-TPFP module. (c) A brief statistic of the previously reported THz lithium niobate sources with THz pulse energy as a function of the operating repetition rate. The dashed line depicts the levels of the different THz output average powers.}
\label{fig1}
\end{figure}

Our high-average-power THz source was powered by a Yb:YAG high-average-power high-repetition-rate industrial laser (Ultron Photonics Technology Co., Ltd., OR-50-IR-A). This laser had a center wavelength of 1030 nm, a typical repetition rate of 50 kHz, a maximum output power of \textasciitilde 50 W, pulse energy of \textasciitilde 1 mJ, a full-width-at-half-maximum (FWHM) spot diameter of \textasciitilde 2 mm, and a Fourier-transform-limited pulse duration of \textasciitilde 570 fs. The repetition rate of this laser can be modulated below 50 kHz upon experiment requirements, and the laser pulse energy remains unchanged. Furthermore, the crystal used was 5 mol\% MgO congruent lithium niobate (YunYi-Tech Co., Ltd.), prism-shaped, 30-mm high, with an isosceles triangle bottom having two 20-mm waists and two 63° base angles. Anti-reflection coatings were applied to the two waist surfaces, and the crystal was operated at room temperature. The overall schematic diagram of this THz source and the detection system is illustrated in Fig.~\ref{fig1}(a) and (b). The 1030-nm pumping laser was separated into two parts. Approximately 95\% of the laser energy was used as the pumping laser, while the remaining was guided to a commercially available optical parametric amplifier to generate a \textasciitilde 800-nm probe laser.

In the high-energy pumping part, after passing through a motorized linear stage, the pumping laser beam was guided by a rail periscope into a grating-TPFP module, which can be replaced by an echelon-TPFP module, depicted in Fig.~\ref{fig1}(b). Subsequently, the pumping laser traversed a dual-lens imaging system and was finally incident onto the crystal. The magnification ratio of the pumping laser can be adjusted by replacing the first mirror in the imaging system. Additionally, an optical chopper was added to the pumping optical path when measuring the THz temporal waveform. After the THz output surface of the lithium niobate crystal, the redundant pumping laser was blocked by a 20-\textmu m black polyimide thin film, with a THz power transmittance of \textasciitilde 0.8. Then, the output THz pulses were collected and focused to a 50-\textmu m-thick gallium phosphide crystal (GaP) by two off-axis parabolic mirrors and an indium tin oxide glass. The first off-axis parabolic mirror had a focal length of 3 inches and a diameter of 3 inches, facilitating the collection of divergent THz waves. The second off-axis parabolic mirror had a focal length of 2 inches and a diameter of 2 inches, focusing the THz pulse into a small spot. In addition, a beam expander comprising a pair of concave and convex lenses can be added before the grating-TPFP module to decrease pumping fluence.

In the remaining low-energy probing part, the output laser from the optical parametric amplifier was compressed by a prism pair to \textasciitilde 50 fs. Then the probe laser beam was delayed by a motorized linear stage, aligned collinearly with the THz beam by the indium tin oxide glass, and concurrently focused on the detection crystal by the second off-axis parabolic. The THz temporal waveforms were measured by conventional electro-optic sampling (EOS). The THz energy and power were detected by a THz power meter (Ophir, 3A-P-THz), while the THz beam profile at the focal point was measured by a THz camera (Ophir, Pyrocam III HR). Additionally, the power of the pumping laser illuminated the lithium niobate crystal was measured after the grating or echelon mirror, and the beam profile of the pumping laser at the input surface of the lithium niobate crystal was measured by a CMOS camera (Basler, acA1600-60gm). It is important to note that both the THz and laser spots were measured at <1~kHz repetition rate to avoid device damage. 
Fig.~\ref{fig1}(c) depicts the summarized statistics of the previously reported THz sources and the current performance of our system, characterized by the pulse energy and repetition rates\cite{guiramandNearoptimalIntensePowerful2022,vogelSinglecycle643MW2024,ofori-okaiTHzGenerationUsing2016,kramerEnablingHighRepetition2020,yehGeneration10MJ2007,hoffmannFiberLaserPumped2008,hiroriSinglecycleTerahertzPulses2011,fulopGenerationSubmJTerahertz2012,huangHighConversionEfficiency2013, vicarioPumpPulseWidth2013,fulopEfficientGenerationTHz2014,schneider800fs330mJPulses2014,ochiYbYAGThindisk2015,wuHighlyEfficientGeneration2018,meyerSinglecycleMHzRepetition2020,zhang14mJHighEnergy2021,millon400KHzRepetition2023,wangHighpowerIntracavitySinglecycle2023,namOptimizedTerahertzPulse2023}. The lithium niobate crystals operated at low temperatures (LT) and room temperatures (RT) are classified into two groups.

\subsection{Grating-TPFP module}

The narrow bandwidth of the pumping laser mitigates the degradation of pulse duration caused by angular dispersion\cite{ofori-okaiTHzGenerationUsing2016}. Therefore, the grating-TPFP technique theoretically performs better using this Yb laser than 30-fs Ti:sapphire lasers. In the grating-TPFP module, the adopted grating had a groove density of 1000 lines/mm. Two mirrors were employed to align the laser to target the transmissive grating at an incident angle of 31°, corresponding to the Littrow angle. Then, another two mirrors reflected the –1st-order diffracted laser to the imaging system. In the imaging system, the first and second lenses (L1 and L2) had a focal length of 370~mm and 100~mm, respectively, producing a reduction ratio of 3.7. The pumping laser transmittance of the grating module was >90\%.

\subsection{Echelon-TPFP module}
An echelon mirror is capable of generating a discrete tilted pulse front, which avoids angular dispersion and increases THz conversion efficiency\cite{ofori-okaiTHzGenerationUsing2016,guiramandNearoptimalIntensePowerful2022}. Therefore, this scheme is a strong competitor for grating-TPFP. In the echelon-TPFP module, as illustrated in Fig.~\ref{fig1}(b), the reflection echelon mirror had a size of 1 inch ×1 inch, with a step width of 150 \textmu m and a step height of 54.6 \textmu m. By introducing the design of an optical isolator\cite{chengOnchipPoorMans2022}, a normal incidence configuration could be achieved for the echelon mirror, which avoided distortion of the laser spot caused by oblique incidence. The perpendicularly polarized (s-polarized) pumping laser was converted to a parallelly polarized (p-polarized) one with the assistance of a half-wave plate. Then, it passed through a polarizing beam splitter (PBS, AOI 45°, the s-polarized laser is reflected while the p-polarized laser is transmitted) and illuminated the echelon mirror in the normal direction. A quarter-wave plate was inserted between the PBS and the echelon mirror. The pumping laser passed through the quarter-wave plate twice and caused a $\pi$ retard between its components. Consequently, the polarization of the reflected laser was reverted to s-polarized, which was then reflected by the PBS and guided to the imaging system. In the imaging system, the focal length of the first lens (L1)  was 300 mm, resulting in a shrink ratio of 3. In order to cover more steps of the echelon mirror, an expander with a magnification ratio of 2 was added between the quarter wave plate and the echelon mirror. This expander comprised two lenses with focal lengths of –50 mm and 100 mm, respectively. It should be noted that the expander also narrowed the pumping beam when it returned, and the total reduction in beam diameter from the echelon mirror to the crystal was designed to be 6. In this design, the expander and the imaging system collectively performed the function of beam compression, making it possible to use higher compression ratios. A further elevation of compression ratios can subsequently reduce the step height of the echelon mirror and ease its manufacturing process. Besides, this new in-plane echelon-TPFP design can provide more space for further improvement and can also be used in single-shot detection systems. The laser transmittance of this echelon-TPFP module was >70\%, and practically all the loss was due to the echelon mirror.

\section{Results and discussion}

\subsection{Grating-TPFP THz source at 1 kHz}

\begin{figure}[htb]
\centering\includegraphics[width=\textwidth]{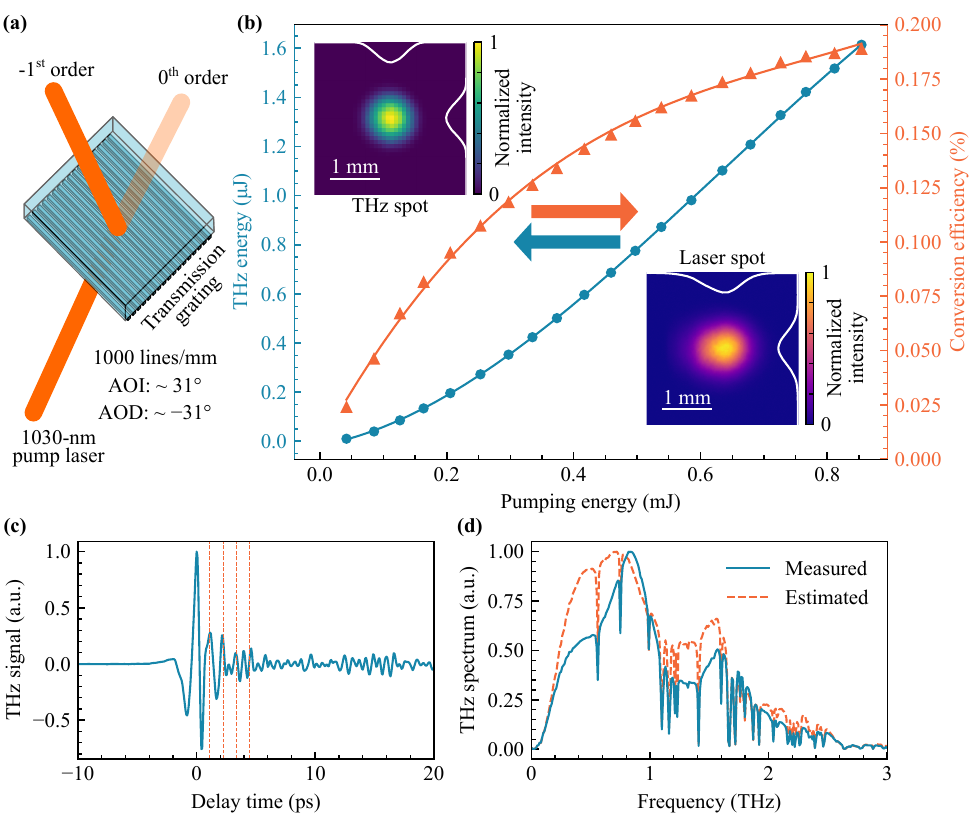}
\caption{The THz characteristics of the strong-field 1-kHz THz source based on the grating-TPFP. (a) The designed angle of incidence (AOI) and dispersion (AOD) for the transmission grating. (b) The THz pulse energy (circle) and optical-to-THz conversion efficiency (triangle) vary with pumping energy, measured at a relative humidity of \textasciitilde 40\%. The points are original data, while the curves are third-order polynomial fitted results. The insets are the THz focal spot (upper left) and the laser spot (lower right) measured at the input surface of the lithium niobate crystal. (c) The normalized THz temporal waveform. The vertical dashed lines show the calculated locations of the first four reflected pulses in the GaP crystal. (d) The corresponding normalized THz spectrum (solid), together with the estimated THz spectrum (dashed) where the influence of the multi-reflection in detection crystal is removed semi-empirically.}
\label{fig2}
\end{figure}

To investigate the THz pulse energy generated by this 1030-nm industrial laser, the grating-TPFP module was constructed, optimized, and measured under a repetition rate of 1 kHz. In this grating-TPFP system, the beam expander in the pumping optical path was omitted to achieve a high pumping fluence. The configurations of the grating are shown in Fig.~\ref{fig2}(a). The output THz pulses were evaluated in three aspects: pulse energy, focused spot, and temporal waveform. 

First, the maximum THz pulse energy obtained from this strong-field grating-TPFP 1-kHz THz source is 1.6 \textmu J, with a pumping pulse energy of 0.85 mJ and a corresponding conversion efficiency of 0.19\%. As depicted in Fig.~\ref{fig2}(b), a third-order polynomial relationship is found for the measured THz-to-laser energy dependence. Second, the focused THz spot in this system has a quasi-Gaussian profile, as shown in the upper left inset of Fig.~\ref{fig2}(b), which has FWHM major × minor axis lengths of 0.64 mm × 0.62 mm, resulting in a remarkable ellipticity of 0.97. The laser spot, measured at the input surface of the lithium niobate crystal, is illustrated in the lower right inset of Fig.~\ref{fig2}(b), which has FWHM major × minor axis lengths of 0.83 mm × 0.70 mm. Third, the THz temporal waveform is shown in Fig.~\ref{fig2}(c), which is measured under an ambient environment and exhibits several minor oscillations caused by water vapor absorption. The corresponding spectrum ranges from 0.1 THz to over 2.5 THz, as shown in Fig.~\ref{fig2}(d). However, there are two obvious dips located at \textasciitilde 0.6 THz and \textasciitilde 1.3 THz due to the THz pulse echoes in the detection crystal. We tried to remove the influence of the echoes on the THz spectrum by dividing it with a thin film transmittance curve\cite{roggenbuckCoherentBroadbandContinuouswave2010,fanFastEnergyDetection2025} (see Appendix~A). The first dip in the estimated spectrum was well relieved, as shown by the dashed curve in Fig.~\ref{fig2}(d). Nevertheless, the second one was not completely removed, possibly due to the imperfect normal incidence of THz pulses on the crystal. Using the measured THz temporal waveform, the energy, and the focal spot profile, the THz peak field at the focal point is calculated to be \textasciitilde 466 kV/cm. The detailed calculation process is shown in Appendix~B.

\subsection{Echelon-TPFP THz source at 1 kHz}

\begin{figure}[htb]
\centering\includegraphics[width=\textwidth]{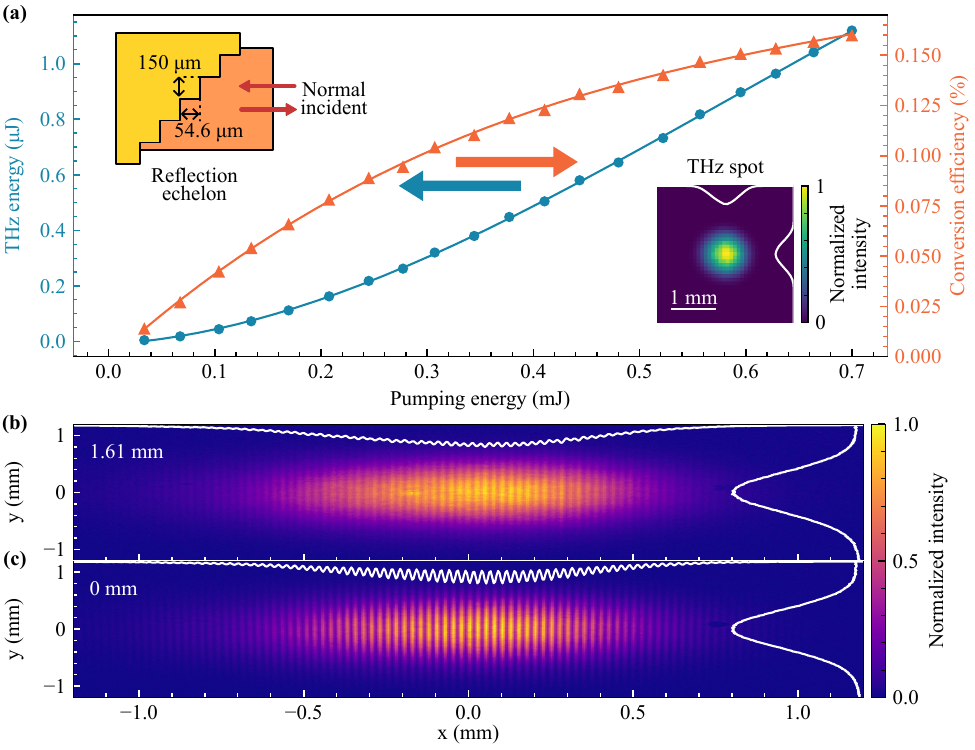}
\caption{The THz characteristics of the strong-field 1-kHz THz source based on the echelon-TPFP. (a) THz pulse energy (circle) and efficiency (triangle) vary with pumping energy, where the points are original data and the curves are third-order polynomial fitted results. The upper left inset shows the normal incidence of the echelon mirror, and the lower right inset shows the THz focal spot. The relative humidity is \textasciitilde 45\% when measured. (b) The laser beam spot at the input surface of the lithium niobate crystal, which is 1.61 mm behind the imaging plane. The x-axis is elongated to better show the minor segments. (c) The laser spot at the imaging plane, determined by its high contrast ratio.}
\label{fig3}
\end{figure}

A comprehensive investigation was also carried out on the echelon-TPFP scheme, to find a more effective implementation scheme for the TPFP technique. The beam expander in the pumping optical path was not installed in this experiment. The THz energy and efficiency both monotonically increase with the pumping energy, as shown in Fig.~\ref{fig3}(a). The trends of these two curves are analogous to those in the grating-TPFP system. However, the maximum pumping energy is limited to 0.7 mJ, due to the low reflection of the echelon mirror. The largest THz pulse energy is 1.1 \textmu J, and the efficiency is 0.16\%. Shown in the lower right inset of Fig.~\ref{fig3}(a), the THz FWHM spot size is 0.66 mm (major) × 0.61 mm (minor), and the ellipticity is 0.94, which is close to that in the grating system. The THz temporal waveform is shown in Fig.~\ref{fig8} (Appendix~C). Using these measured results, the calculated peak electric field is 422 kV/cm, which is lower than that in the grating system. 

The laser beam spot at the input surface of the lithium niobate crystal shows a quasi-Gaussian profile divided into multiple segments arranged in the horizontal direction, as shown in Fig.~\ref{fig3}(b). This spot has FWHM axis lengths of around 0.88 mm × 0.82 mm, which is also close to that observed in the grating-TPFP system. The beam spot covers >60 steps of the echelon mirror, producing a dense, discrete, tilted pulse front. Moreover, the crystal input surface is 1.61 mm behind the imaging plane and the laser spot measured at the input surface has a lower contrast ratio than that at the imaging plane, shown in Fig.~\ref{fig3}(c), which is caused by diffraction and imperfect imaging. We anticipate that an almost continuous pulse front, resembling the one generated by a grating, could be achieved by further increasing echelon mirror steps illuminated by the pumping laser. Instead of reducing individual step width to increase step density, the laser spot can be enlarged by the expander in the echelon-TPFP module.

\subsection{High-average-power 50-kHz THz source}
We chose the grating-TPFP design to achieve a high-average-power THz source. The echelon-TPFP generates lower THz energy than the grating-TPFP, which is, on one hand, due to its low transmission. On the other hand, this may be because the pumping laser has a narrow spectral range, where the angular dispersion of the grating is reduced, suppressing the advantages of the echelon mirror. 

In the aforementioned grating-TPFP system, the repetition rate of the pumping laser was gradually increased to 50 kHz. However, the lithium niobate crystal broke at 25 kHz, with its damage threshold located between 2.4~$\mathrm{kW/cm^2}$ and 3.2~$\mathrm{kW/cm^2}$ (the highest power). The possible cause of this fracture is the light-induced absorption inside the lithium niobate crystal\cite{bachLaserInducedDamage2017}. To mitigate the light-induced absorption and circumvent the crystal disruption, an expander with a designed ratio of 1.5 is added to the pumping optical path, shown in Fig.~\ref{fig1} (near the optical parametric amplifier). Finally, through expansion of the pumping laser and system optimization under a \textasciitilde 13\% relative humidity, we successfully operated the modified grating-TPFP system at 50-kHz repetition rate. 

\begin{figure}[htb]
\centering\includegraphics[width=\textwidth]{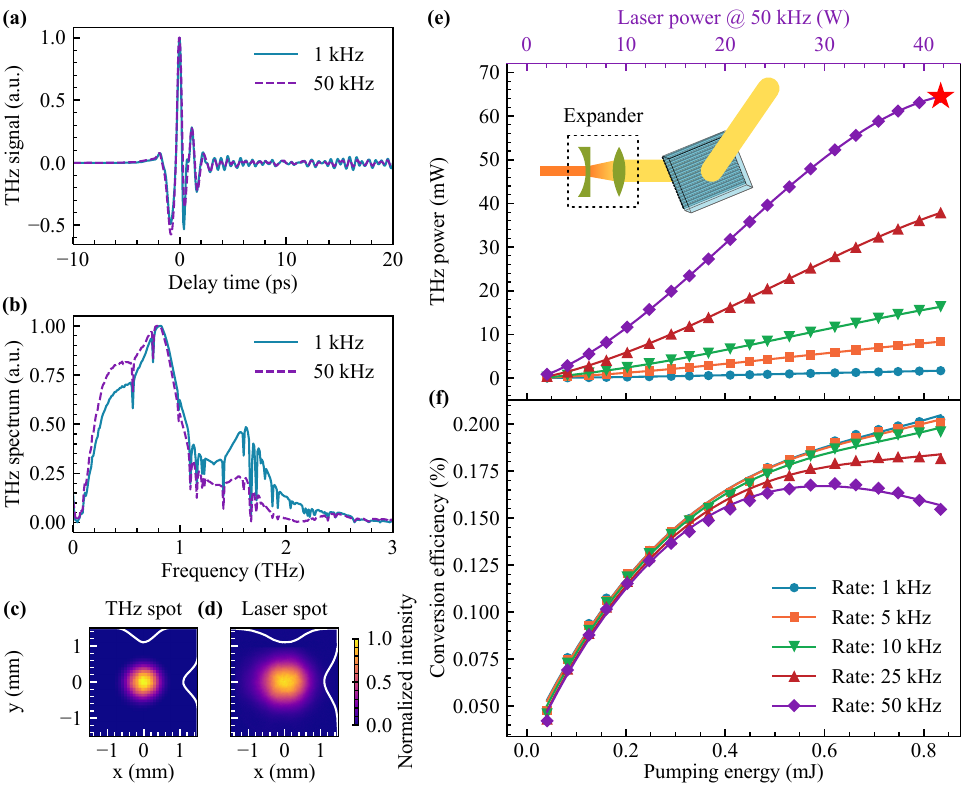}
\caption{The THz characteristics of the high-average-power 50-kHz THz source based on the grating-TPFP with an expanded pumping laser. (a) Normalized THz temporal waveforms at 1-kHz and 50-kHz repetition rates, respectively. (b) Corresponding normalized THz spectra. (c) THz focal spot and (d) laser beam spot measured at the input surface of the lithium niobate crystal. (e) The THz power curves and (f) corresponding efficiency curves measured at different repetition rates. The points are original data and the curves are third-order polynomial fitted results. The top purple axis shows the pumping power at 50-kHz repetition rate, calculated by multiplying pumping energy by repetition rate. All the THz data in this figure is measured at a relative humidity of \textasciitilde 13\%.}
\label{fig4}
\end{figure}

To achieve an elevated THz pulse energy and peak electric field, the modified system was characterized at a low relative humidity of \textasciitilde 13\%, where the water vapor absorption was low. The THz temporal waveforms and the corresponding spectra at 1-kHz and 50-kHz repetition rates are shown in Fig.~\ref{fig4}(a) and (b), respectively. In order to accurately measure the THz temporal waveforms at high pumping power, the chopper was positioned in the detection optical path. At a 50-kHz repetition rate, the high-frequency components of the THz pulse are partially lost, resulting in a slight pulse broadening. The THz focal spot shown in Fig.~\ref{fig4}(c) has FWHM major × minor axis lengths of 0.66 mm × 0.62 mm, with an ellipticity of 0.95. The axis lengths of the laser beam spot are 1.04 mm × 0.92 mm, with an ellipticity of 0.88, as shown in Fig.~\ref{fig4}(d). The area of the laser spot at the input surface of the lithium niobate crystal is \textasciitilde 1.6 times larger than that without the expander. Fig.~\ref{fig4}(e) and (f) show that the THz energy and conversion efficiency vary with pumping energy at different repetition rates. At a 1-kHz repetition rate, the maximum THz pulse energy is 1.7 \textmu J with a pumping laser energy of 0.83 mJ, and the corresponding efficiency is 0.20\%. The calculated peak electric field is 525 kV/cm. As shown in Fig.~\ref{fig4}(f), the conversion efficiency at high pumping energy decreases when the repetition rate increases. Our system can produce a THz average power of 64.5 mW at a repetition rate of 50 kHz, which approaches the leading level worldwide. Additionally, the corresponding THz pulse energy and conversion efficiency are 1.3 \textmu J and 0.15\%, respectively. The peak electric field is calculated to be 426 kV/cm. At a high repetition rate, heat is a major hindrance to THz conversion efficiency. Therefore, it is necessary to export the heat inside the crystal using a suitable method to improve THz output further. 

\begin{figure}[htb]
\centering\includegraphics[width=\textwidth]{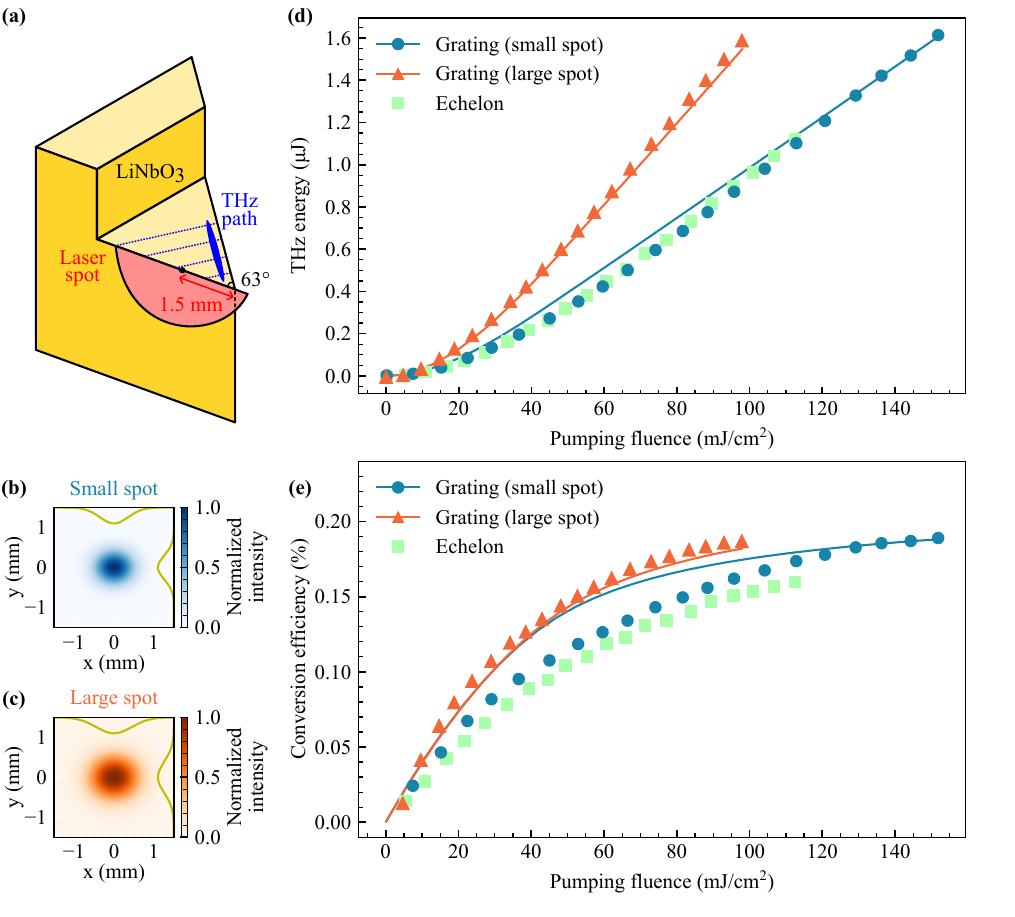}
\caption{Influence of pumping fluence and Gaussian spot profile on THz efficiency. (a) The schematic diagram and configuration of the model. The center of the pumping laser spot is 1.5 mm from the crystal edge. The angle of the crystal is 63°. (b) and (c) are the small and large input Gaussian spots for the two grating systems without/with the expander, respectively. (d) The THz energy curves and (e) conversion efficiency curves for grating systems with small/large spots and the echelon system. The points are measured results, and the curves are simulated results.}
\label{fig5}
\end{figure}

In order to investigate the effect of the pumping beam size on the THz generation efficiency, we also measured the THz energy of this high-average-power system operating at \textasciitilde 40\% relative humidity, as illustrated in Fig.~\ref{fig9} (Appendix~D). The maximum THz pulse energy reaches 1.6~\textmu J with a conversion efficiency of 0.19\%, which is identical to that without the expansion of the pumping laser beam. It is unusual that, despite a reduction of \textasciitilde 1.6 times in the pumping fluence via expanding, the maximum THz conversion efficiency and the corresponding THz pulse energy show minimal variation. To analyze the underlying mechanisms of this phenomenon, a simple empirical model is introduced. This model, as shown in Fig.~\ref{fig5}(a), includes the following characteristics: (1) the THz conversion efficiency varies with the localized pumping fluence and saturates at the peak point; (2) the THz waves are assumed to be instantaneously generated when the pumping laser contacts lithium niobate crystal, and the generated THz intensity distribution is calculated by the efficiency curve; (3) the absorption of THz waves is estimated to be 1.5~mm$\mathrm{^{-1}}$ \cite{unferdorbenMeasurementRefractiveIndex2015}; (4) the distance is estimated to be 1.5~mm between the center of the pumping laser spot and the crystal edge; (5) the tilted pulse front angle is 63°; (6) the input pumping spot is Gaussian. The efficiency $v$ is calculated by an empirical piecewise function (Eq. \ref{eq1}).

\begin{equation}
v=\left\{\begin{matrix} 
   V_{sa}\left ( 1- \left ( 1-\frac{F}{F_{sa}} \right ) ^2  \right ) & F<F_{sa}  \\
   
   V_{sa} & F \ge F_{sa}
\end{matrix}\right. 
\label{eq1}
\end{equation}

\noindent where $V_{sa}=1.4\%$ is the saturated efficiency, $F$ is the localized pumping fluence, and $F_{sa}=40~ \mathrm{mJ/cm^2}$ is the saturated pumping fluence. The input pumping spot sizes are calculated by the fitting results of the measured spots, as shown in Fig.~\ref{fig5}(b) and (c). The simulated results are shown in Fig.~\ref{fig5}(d) and (e), together with the measured data in three systems: grating-TPFP system with/without the beam spot expansion, and echelon-TPFP system. The calculated THz energy curves agree well with the measured data. The mismatch of the efficiency curve for the small spot condition in Fig.~\ref{fig5}(e) may be due to the imperfect empirical efficiency curve, along with the neglect of factors such as the cascade effect and the effective generation length that affect THz efficiency. This model qualitatively shows that when we use a Gaussian beam to generate THz pulses, the area near the peak point may be already saturated even though the efficiency curve is not saturated. Close to the pumping fluence that causes saturation, the THz efficiency may not decrease even if the pumping laser spot is enlarged. 

\subsection{THz induced magnetization dynamics in CoFe thin film}

\begin{figure}[htb]
\centering\includegraphics[width=\textwidth]{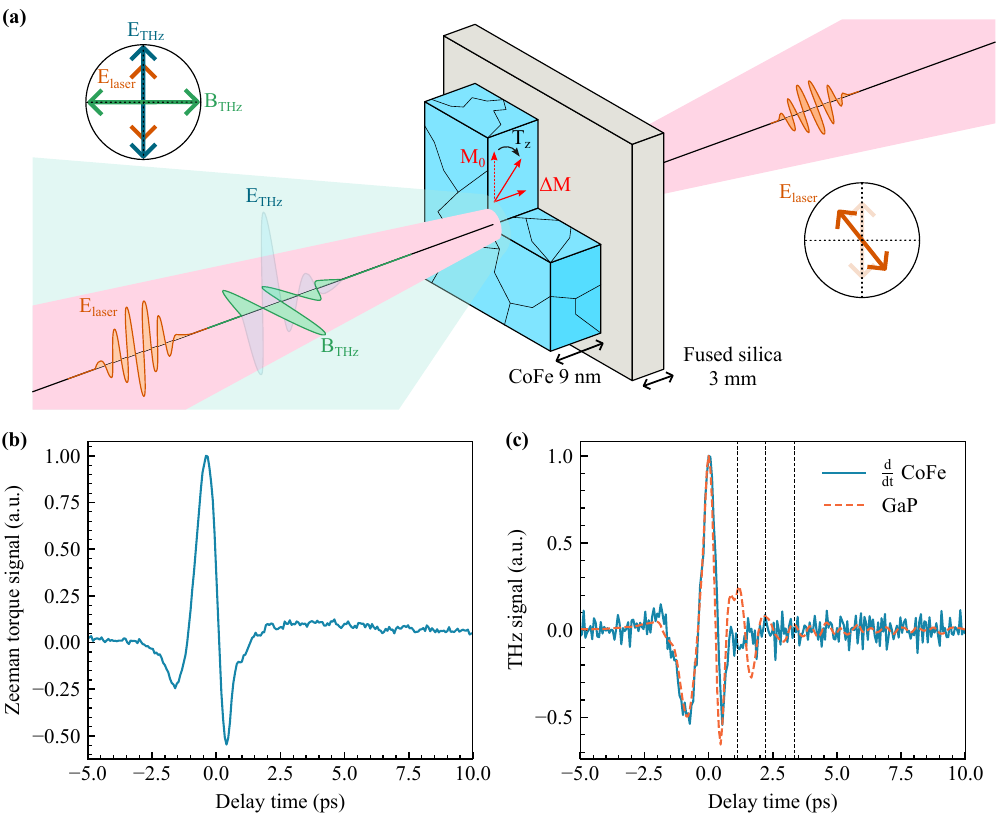}
\caption{Magnetization dynamics in CoFe thin film induced by strong-field THz pulse. (a) Schematic diagram of THz pump-probe experiment and the polarization configurations of THz and laser pulses. (b) The normalized Zeeman torque signal. (c) The differential and normalized signal (solid) of the original signal in (b), compared with the EOS signal (dashed) measured by the GaP crystal. The vertical dashed lines show the calculated locations of the first three reflected pulses.}
\label{fig6}
\end{figure}

The high THz magnetic fields enable investigations of magnetization dynamics in ferromagnetic materials. A ferromagnetic thin-film material, consisting of the 9-nm CoFe magnetron sputtered on a 3-mm thick fused silica glass, was pumped by the THz pulses and measured by the probe laser in our high-average-power system at a repetition rate of 50 kHz. 

As shown in Fig.~\ref{fig6}(a), an external static magnetic field $M_0$ (dotted arrow) is applied to the thin film. In the ferromagnetic material, the strong THz magnetic field exerts a torque $T_Z$ that swivels the dynamic magnetization out of the plane, and the probe laser can measure this out-of-plane component $\Delta M$ \cite{chekhovBroadbandSpintronicDetection2023}. The measured Zeeman torque signal, as shown in Fig.~\ref{fig6}(b), shows a single-cycle THz pulse followed by minor oscillations derived from water vapor absorption. After acquiring the Zeeman torque signal, the THz temporal waveform can be extracted by its time derivative. In the extracted THz signal, shown in Fig.~\ref{fig6}(c), no echoes are observed, which can be found in the EOS signal. The echoes are postponed out of the time window by the 3-mm thick substrate. The successful acquisition of the Zeeman torque signal demonstrates the capability of our system to measure such faint signals. However, the signal-to-noise ratio is lower for the extracted THz temporal waveform from the Zeeman torque signal than that from EOS, even though the time constant of the lock-in amplifier is set to \textasciitilde 5 s and the measurement time reaches \textasciitilde 42 s for each point. We have not yet fully exploited the high THz average power because of the substantial difference between the chopping frequency (<10~kHz) and the laser repetition rate (50~kHz), and the detection method requires further study. 

\section{Conclusion}

In summary, a high-average-power THz source is demonstrated by using a 50-W 1030-nm Yb:YAG femtosecond laser with a pulse duration of \textasciitilde 570 fs, a pulse energy of \textasciitilde 1 mJ, and a repetition rate of 50 kHz. Two competing TPFP implementations (the grating-TPFP and the echelon-TPFP designs) are constructed and systematically compared, and the grating-TPFP design is more suitable for our system. At a lower relative humidity of \textasciitilde 13\%, utilizing a pumping laser energy of 0.83 mJ, we achieved a THz pulse energy of 1.7 \textmu J at a 1-kHz repetition rate. The corresponding conversion efficiency is 0.20\%, and the calculated peak electric field is 525 kV/cm. At a 50-kHz repetition rate, we achieve an outstanding THz output of 64.5 mW with a peak electric field of 426 kV/cm, which approaches the global leaders. Our system demonstrates that high-repetition-rate strong-field THz pulses can be directly generated using industrial Yb-doped lasers with pulse durations typically exceeding 500 fs, thereby eliminating the necessity for additional pulse compression techniques. This scheme greatly reduces the overall cost of strong-field THz systems, thereby facilitating the widespread adoption of strong-field THz sources. Besides, we successfully achieved the Zeeman torque signal in the CoFe thin film, demonstrating the measurement ability of our system. Our high-average-power THz system has the potential to promote the development of weak signal detection, THz spectroscopy, THz imaging, and other applications sensitive to high repetition rates and measurement time.

\appendix

\section*{Appendix A: The removal of echoes in the GaP crystal }
\label{Appx1}

\begin{figure}[htbp]
\centering
\includegraphics[width=.35\linewidth]{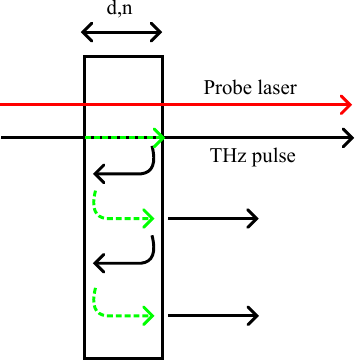}
\caption{Reflection and transmission of THz pulses in a thin film.}
\label{fig7}
\end{figure}

In the detection crystal of an EOS system, only the THz pulses that propagate in the same direction as the probe laser can be detected, shown as the dashed arrows in Fig.~\ref{fig7}. Fortunately, since each of these pulses will eventually pass through the output surface once, the total transmittance curve of the thin film can be used to calculate the pulses that can be measured. Moreover, because the GaP crystal has a thin thickness of 50~\textmu m, the number of reflections can be assumed to be infinite.

The transmittance of a thin film is calculated by Eq. \ref{eqA1}\cite{roggenbuckCoherentBroadbandContinuouswave2010,fanFastEnergyDetection2025}.

\begin{equation}
T=\frac{\left ( 1-R \right ) ^2 }{\left ( 1-R \right ) ^2+4R \sin^2{\left ( n\omega d / c \right ) }} 
\label{eqA1}
\end{equation}

\noindent where $R=\frac{\left ( n-1 \right )^2 }{\left ( n+1 \right )^2} $,  $n$ is the refractive index of GaP at THz frequency range\cite{wu7TerahertzBroadband1997}, $\omega$ is the angular frequency, $d$ is the thickness of GaP, $c$ is the speed of light in air. The influence of the echoes is mitigated by dividing the spectrum with the thin film transmission ratio and then the curve is normalized.

\section*{Appendix B: Extract the peak electric field of a THz pulse }
\label{Appx2}

The peak electric field of a THz pulse can be obtained by Eq. \ref{eqA2}\cite{zhang14mJHighEnergy2021}.

\begin{equation}
E=\sqrt{\frac{\mathcal{E} }{c \varepsilon_0 \tau A } }
\label{eqA2}
\end{equation}

\noindent where $E$ is the peak electric field, $\mathcal{E}$ is the THz pulse energy, $c$ is the speed of light in air, $\varepsilon_0$ is the vacuum permittivity, $\tau=\int_{t} \left | E\left ( t \right )  \right |^2 dt $ is the time parameter, where $E\left ( t \right ) $ is the normalized THz temporal profile and $A=\int_{t} \left | E\left ( x,y \right )  \right |^2 dx dy$ is the space parameter, where $E\left ( x,y \right ) $ is the normalized spatial profile. In our method, $\tau$ can be calculated from the normalized THz waveform, and $A$ can be evaluated from the integral of the measured and normalized THz spot profile (summation of discrete values in THz spot data and then multiply the summation by unit area), as $I\propto \left | E\left ( x,y \right )  \right |^2 $.

\section*{Appendix C: The THz temporal waveform for the echelon-TPFP system}
 \label{Appx3}
 
\begin{figure}[htbp]
\centering
\includegraphics[width=\textwidth]{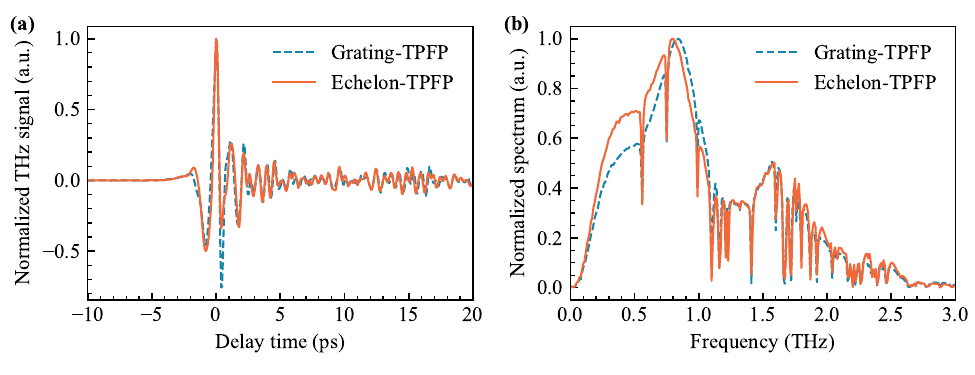}
\caption{The THz temporal waveforms. (a) The waveforms for grating-TPFP and echelon-TPFP systems. (b) The corresponding THz spectra.}
\label{fig8}
\end{figure}

As shown in Fig.~\ref{fig8}(a), the THz temporal waveforms of the grating-TPFP and echelon-TPFP systems are similar. The difference may be due to the variation of the THz focusing and system optimization. 

\section*{Appendix D: The THz output energy of the high-average-power system at the relative humidity of ~40\% }
\label{Appx4}

\begin{figure}[htbp]
\centering
\includegraphics[width=\textwidth]{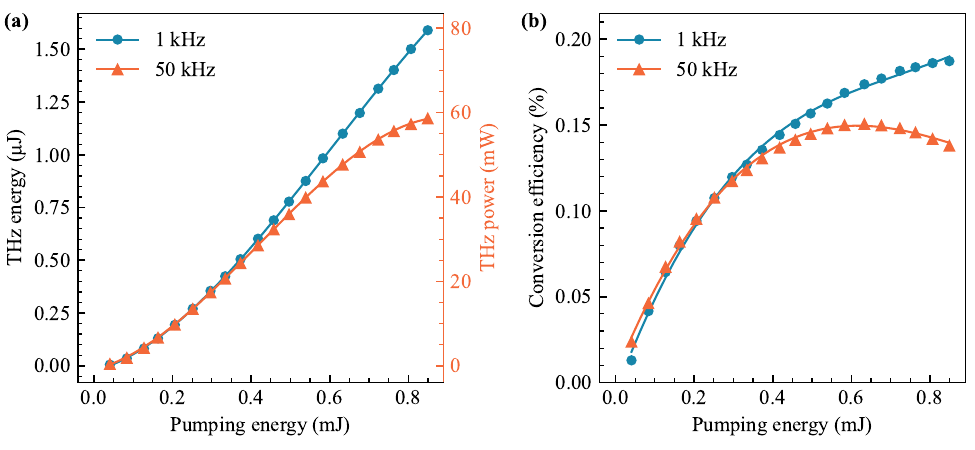}
\caption{The THz output of the high-average-power system at the relative humidity of ~40\% (a) The THz energy curves and (b) the THz conversion efficiency curves at the repetition rates of 1 kHz and 50 kHz, respectively. The right y-axis in (a) shows the corresponding THz power at the repetition rate of 50 kHz.}
\label{fig9}
\end{figure}

The THz energy and conversion efficiency of the high-average-power system at the relative humidity of ~40\% is shown in Fig.~\ref{fig9}. At the repetition rate of 1 kHz, the maximum THz pulse reaches 1.6 \textmu J with a conversion efficiency of 0.19\%. At the repetition rate of 50 kHz, the highest THz power is 58.7 mW, and the corresponding efficiency is 0.14\%.

\begin{backmatter}
\bmsection{Funding}
This work was supported by the National Natural Science Foundation of China (U23A6002, 92250307), the National Key R\&D Program of China (2022YFA1604402), the Open Project Program of Wuhan National Laboratory for Optoelectronics NO.2022WNLOKF006. We are also grateful for financial support from the Youth Beijing Scholar Sponsorship Program.

\bmsection{Acknowledgment}
We thank LBTEK for providing optical components and mechanics.

\bmsection{Disclosures}
The authors declare no conflicts of interest.

\bmsection{Data Availability Statement}
The data that support the plots within this paper and other findings of this study are available from the corresponding author upon reasonable request.

\end{backmatter}


\bibliography{refs}

\end{document}